\documentclass[12pt, a4paper]{article}

\pdfoutput=1

\usepackage{amsmath}
\usepackage{amsfonts}
\usepackage{amssymb}
\usepackage{cite}
\usepackage{multicol,multirow}
\usepackage{graphicx, physics, subcaption}
\usepackage{graphics, epsfig}
\usepackage{epsf}
\usepackage{epstopdf}
\usepackage[usenames,dvipsnames]{color}

\usepackage{colortbl}
\definecolor{summersky}{cmyk}{0.71,0.33,0,0.14}
\definecolor{flamingo}{cmyk}{0,0.51,0.71,0.14}
\definecolor{rp}{cmyk}{0.2, 1, 0.6, 0}
\definecolor{pacificblue}{cmyk}{0.95,0.3,0, 0.19}
\definecolor{gray60}{cmyk}{0.4,0.4,0,0.8}

\numberwithin{equation}{section}

\usepackage[colorlinks,bookmarks=false, hypertexnames=true]{hyperref}
\hypersetup{
	breaklinks=true,
	pdfstartview={FitH},    
	colorlinks=true,       
	linkcolor=blue,          
	citecolor=red,        
	filecolor=magenta,      
	urlcolor=red,           
	anchorcolor=green,      
	linktocpage=true
}

\newcommand{\nc}{\newcommand}

\nc{\ba}{\begin{eqnarray}}
\nc{\ea}{\end{eqnarray}}

\nc{\calR}{{\cal{R}}}
\nc{\calP}{{\cal{P}}}
\nc{\cN}{ {\cal{N}} }

\def\lmax{\ell_{\mathrm{max}}}

\begin{document}

\def\thefootnote{\fnsymbol{footnote}}

\begin{center}

{\bf  Bound States Inside the Horizon and  Black Hole Entropy }
\\[0.5cm]

{
Hassan Firouzjahi\footnote{firouz@ipm.ir}
}
\\[0.5cm]
 
 {\small \textit{School of Astronomy, Institute for Research in Fundamental Sciences (IPM) \\ P.~O.~Box 19395-5746, Tehran, Iran
}}\\

\end{center}

\vspace{.3cm}
\hrule
\begin{abstract}

We study  the bound states perturbations  in the interior of 
the Schwarzschild black hole and examine whether or not their microstates spectrum  can furnish  a statistical description of the black hole entropy.   These states contain a discrete set of  modes characterized by the labels $(s, p,\ell,m)$, where $s$ represents the spin of the species, 
$p$ denotes the radial excitation number while $(\ell,m)$ are the usual angular momentum labels.   Requiring that the shortest wavelength resolved by the highly excited bound states remains above the 
quantum gravity scale introduces an angular cutoff $\ell_{\rm max}\sim
r_H/L_{\rm Pl}$. The resulting angular degeneracy therefore scales
with the horizon area. We count the corresponding configurations in the microcanonical ensemble and find an entropy of the form $S_{\rm bs}=\alpha A/\hbar G$. Demanding that the bound state sector to be in thermal equilibrium with the black hole at the Hawking temperature fixes the fraction of the bound state energy in total black hole energy.  For the gravitational perturbations sector, the equilibrium bound state contribution in entropy and the energy fraction  are typically small.   We also
examine the conditions under which the bound states could account for
the full  black hole entropy. 
Although such saturation is possible at the level of state counting, it requires an order unity fraction in the black hole energy and therefore the backreactions on the background geometry cannot be ignored. \\

\end{abstract}

\vspace{0.3cm}
\hrule

\newpage

\section{Introduction}
\label{sec:intro}
The discovery of the black hole (BH) thermodynamics represents a deep connection between gravity, quantum theory, and statistical mechanics. The Bekenstein-Hawking entropy formula\footnote{We work in the unit where  the Boltzmann constant and the speed of light are set to unity.}\cite{Bekenstein1973, Bardeen:1973gs, Hawking1975}
$S_{\rm BH}=\frac{A}{4\hbar G}$
suggests that the entropy of a BH  is proportional to the horizon area $A$ rather than the volume enclosed by the event horizon. Understanding the microscopic origin of this entropy formula is one of the central question in quantum gravity. Several approaches have tried to identify the microscopic degrees of freedom responsible for the BH  entropy \cite{tHooft:1984kcu, Wald:1993nt, Jacobson:1993vj, tHooft:1993dmi, StromingerVafa1996,Rovelli1996,AshtekarBaezCorichiKrasnov1998,BombelliKoulLeeSorkin1986, Srednicki:1993im, Maldacena1998, Padmanabhan:2009vy, Verlinde:2010hp}, such as  the  string theory microstate counting, loop quantum gravity, holographic descriptions, entanglement entropy, Euclidean quantum gravity etc. Despite these progresses, a complete microscopic description of the entropy of 
the  Schwarzschild BH remains an open problem, for a review see 
\cite{Jacobson:1999mi, Wald:2025nbz, Bigatti:1999dp, Solodukhin:2011gn, Elizalde:2025iku}

In this work we explore a different approach. We examine whether or not the bound states in the BH interior can represent 
the corresponding microscopic degrees of freedom to furnish all or a fraction of  the BH entropy. The bound states are the solutions in the interior of BH with the imaginary spectrum  such that they are regular at the center while their profiles fall off exponentially near the horizon so they do not leak to the exterior  region \cite{Firouzjahi:2025jja, Firouzjahi:2025xjp, Steinhauer:2025bbs}.  To be specific, consider the gravitational perturbations in the BH backgrounds. As is well-known, the gravitational perturbations are decomposed into the odd parity (axial) and even parity (polar) perturbations \cite{Chandrasekhar1983}. They are governed by the 
Regge--Wheeler (RW) and Zerilli (Z) master equation respectively   \cite{ReggeWheeler1957,Zerilli1970,Chandrasekhar1983}. The spectrum of these perturbations are vastly studied in the context of quasi normal modes (QNMs) in the exterior region, for a review see \cite{ Berti:2009kk, Konoplya:2011qq, Berti:2025hly}. The QNMs in the exterior region are subject to the incoming boundary condition near the horizon and the outgoing boundary condition on far distances. In this view, the interior bound states  are quite distinct from the exterior QNMs solutions. 
These bound state modes are characterized by a radial excitation number $p$ and angular quantum numbers $(\ell,m)$ and form a finite spectrum for each angular momentum sector. As shown in
\cite{Firouzjahi:2025jja}, for a given value of $\ell$, the axial (RW) perturbations  admit $\ell-2$ bound states. On the other hand, it is shown in \cite{Firouzjahi:2025xjp} that the polar (Z) sector admits $\ell-1$ bound states in which $\ell-2$ of these bound states have the same energy spectrum as the axial sector. As a result, the isospectrality between the axial and polar perturbations for the bound states is satisfied as well. However, it turns out that the additional spectrum in the polar sector is the algebraically special mode (ASM).  

As shown in \cite{Firouzjahi:2025jja, Firouzjahi:2025xjp} the bound state solutions are not restricted to gravitational perturbations. Indeed, other fields of spin $s$ support bound states in the interior of BH. Furthermore, for all spin  $s=0, 1/2, 1, 2$ the highly excited states for large $\ell$ are equally spaced with the spectrum $r_H \omega_I\simeq {p}$ in which $p=1,2,...$, $\omega_I$ is the imaginary part of the spectrum and  $r_H$ is the BH radius. This universal linear spectrum plays important roles in our analysis of identifying the bound states as  possible candidates for the microscopic origin of the BH entropy. Note that  this relation, independent of the spin, identifies the microscopic energy scale of the bound states with the inverse radius of the BH.

With the above discussions in mind, one may ask why the interior bound states should be considered as the microstates furnishing the BH entropy?
Our motivations are as follows. First, the fact that the bound state spectrum are  discrete with labels  $(p, \ell, m)$ is intriguing. Indeed, as we shall see, for $\ell \gg 1$, these labeling introduce a huge angular degeneracy which essentially scales with the horizon area $A$. Second, the universal equally spaced spectrum for excited states leads to a quantized area spectrum \cite{Firouzjahi:2025xjp}.  To see this, suppose we have a transition from two  nearby excited states  with $r_H \Delta \omega_I= 1$. Consider the semi-classical limit where the shift in the spectrum corresponds to a shift in energy  $\Delta E= \hbar \Delta \omega$. This induces a change in BH 
mass $\Delta M= \hbar \Delta \omega_I= \hbar/r_H $. Consequently, this also induces a change in the area as  $\Delta A= 32 \pi G^2 M \Delta M= 16\pi L_{\mathrm{Pl}}^2$ in which  $L_{\mathrm{Pl}}= \sqrt{\hbar G}$ is the Planck length. This provides an additional indication that the bound states spectrum contains information about the quantum structure of the horizon and may be considered as a candidate for the entropy microstates. 
For other proposals in literature yielding to area quantization  see \cite{Bekenstein:1974jk, Bekenstein:1995ju, Mukhanov:1986me, Hod:1998vk, Maggiore:2007nq, Coates:2021dlg}.


\section{ Bound States Inside the  Black Hole}

\label{bound-sec}

We consider the bound states in the interior of the Schwarzschild metric. The background line element is given by, 
\ba
ds^2=
-\Big(1-\frac{r_H}{r}\Big)dt^2 +\Big(1-\frac{r_H}{r}\Big)^{-1}dr^2
+r^2d\Omega^2 ,
\ea
where $r_H=2GM$ is the horizon radius and $d\Omega^2$ represents 
the metric of a unit two-sphere.  Outside the horizon, the coordinate $t$ is timelike and $r$ is spacelike. Inside the horizon their roles are switched and the radial coordinate becomes timelike. This change in the causal structure is essential for the existence of the interior bound states considered in our analysis. 

To set the stage and present the spectrum of bound states, here we review the case of the axial perturbations governed by the RW equation. As we shall see, the main results of the bound states spectrum are shared by all other perturbations including the polar sectors and other spin fields  as well.  There are noticeable differences concerning the total number of bound states in each sector but these differences do not affect our microstate counting as we will elaborate later on.

After expanding the perturbations in spherical harmonics and imposing the 
separation of variables $Z(t, r) \propto e^{- i \omega t} Z(r)$,  the wave equation  can be written in a Schrodinger-like form,
\ba
\frac{d^2 Z}{dr_*^2} + \left(\omega^2-V_\ell(r)\right)Z=0 ,
\ea
where $r_*$ is the tortoise coordinate related to the radial coordinate via,
\ba
\label{r-star}
\dd r_*= \frac{\dd r}{ \big( 1- \frac{r_H}{r}\big)} \, ,
\ea
and $V_\ell(r)$ is the corresponding effective potential. Note that the above differential relation between $r_*$ and $r$ holds for both interior and exterior regions. However, since the roles of $r$ and $t$ are switched in the interior region, the coordinate $r_*$ plays the role of the time 
in the interior of BH. Considering the interior region, the relation between $r_*$ and $r$ is given by,
\ba
\label{r-star-sol}
r_*= r + r_H\ln(1-\frac{r}{r_H}) \, , \quad \quad (r < r_H) \, .
\ea 
Correspondingly, the horizon $r=r_H$ is mapped to $r_* \rightarrow -\infty$ while the center $r=0$ corresponds to  $r_*=0$.

The potential $V_\ell(r)$  inside the BH 
has the same algebraic  form as the standard RW potential  in the exterior region. However, the key conceptual difference is that it describes a time-dependent wave equation instead of  the usual Schrodinger-like wave equation. In addition, the structure of the singularities of the potential near the center is quite distinct. Specifically,  near the center the RW potential scales like $V_\ell(r) \sim 3/r^4 \simeq 3/4 r_*^2$. Because of the non-trivial shape of the potential, the bound state solutions  $\omega= i \omega_I$ with $\omega_I>0$ exist \cite{Steinhauer:2025bbs, Firouzjahi:2025jja, Firouzjahi:2025xjp}. A curious property of these bound state solutions is that their total wavefunction $Z(t, r)$ has a non-zero amplitude near the event horizon \cite{Steinhauer:2025bbs}. 
 
The spectrum of bound states are investigated in  \cite{ Firouzjahi:2025jja, Firouzjahi:2025xjp}. It was shown that for a given value of $\ell$, 
the axial  sector has $\ell-2$ bound states.  The general bound state spectrum has the labels $(p, \ell, m)$ in which $(\ell, m)$ are the usual  angular momentum indices while the index $p$ represents the radial excitations with 
$p=1,2,..., \ell-2$ with $\ell\ge 3$. The radial profile of the RW bound states  with label $p$  has exactly $\ell-p-2$ nodes in the interval $0< r< r_H$. We comment that the highly excited states have lower values of $p$ and hence containing more nodes. On the other hand, the low-lying modes have larger values of  $p$ and hence have fewer nodes. For example, the most excited state corresponds to $p=1$, the next most excited state corresponds to $p=2$ and so on. On the other hand, the lowest state corresponds to $p=\ell-2$ which has no node.  This reverse ordering (excited states having small $p$, low-lying modes having large $p$) may look confusing. But the point is that the actual energy is $\omega^2<0$, so the low-lying states are more negative while the excited states are less negative and hence the reverse  ordering of $p$.  Finally, all bound states are normalizable in the interior of BH.

It was demonstrated in \cite{ Firouzjahi:2025jja} that the spectrum of 
 highly excited states for $\ell \gg 1$ are equally spaced with,
 \ba
 \label{omega-excited}
r_H\,  \omega_I = p + {\cal O}(\ell^{-1}) \, , \qquad \qquad
   (\rm{excited ~ states}, ~   \ell \gg 1). 
 \ea
The above relation is valid to a  very good approximation  for the most excited states, $p=1,2,...$ but the errors become noticeable for large values of $p$. Looking at the structure of the mode function solutions,  one can estimate the maximum value of $p$ beyond which the approximate linear relation  (\ref{omega-excited}) is violated, obtaining $p_{\mathrm{max}}= b \sqrt{\ell}$ 
in which $b$ is a numerical factor of order unity.  
Correspondingly, for $p=1,2,..., p_{\mathrm{max}}$ the linear relation  (\ref{omega-excited}) is valid to good accuracies, the larger is $\ell$ the higher is the accuracy. In table \ref{table-large-ell} we have presented the spectrum associated to highly excited states $p=1,2,3,4$ for the gravitational perturbations with $\ell \gg1$.

\begin{table*}
\centering
\scalebox{0.95}{
\begin{tabular}{|c||c|c|c|c|c|c|c|c|c|}
\hline
$p$ & $\ell=100$ & $\ell=110$ & $\ell=120$ & $\ell=130$ & $\ell=140$ & $\ell=150$ & $\ell=160$ & $\ell=170$ & $\ell=180$\tabularnewline
\hline
1 & 1.0096 & 1.0086 & 1.0079 & 1.0073 & 1.0068 & 1.0063 & 1.0059&1.0056&1.0052 \tabularnewline
\hline
2 & $2.0388$ & 2.0351 & 2.0321 & 2.0296   &  2.0274    &   2.0256 &2.0239&2.0225&2.0212 \tabularnewline
\hline
3 & $3.0889$ & $3.0804$ & 3.0734 & 3.0676&3.0626  & 3.0582 & 3.0545 &3.0512&3.0482  \tabularnewline
\hline
4 & $4.1612$ & $4.1456$ & $4.1327$ & 4.1219 & 4.1127 & 4.1049&4.0980& 4.0920&4.0867\tabularnewline
\hline
\end{tabular}
}
\vspace{0.5cm}
\caption{$ r_H \omega_I$ for $\ell \gg 1$ for the gravitational excited states, $p=1,2,3,4$.   For  a given value of $\ell$, the most excited state is $p=1$, the next most excited state is $p=2$ and so on. The linear relation 
$\omega_ I r_H =p$ is valid to high accuracy for small values of $p$ while 
it is violated for $p\ge p_{\rm{max}}$ with $p_{\rm{max}} \sim \ell^{1/2}$.  }
\label{table-large-ell}
\vspace{.5cm}
\end{table*}

On the other hand, for low-lying modes, corresponding to $p= \ell-2, \ell-3,...$ the spectrum of the bound states can be obtained via a WKB approximation yielding  $r_H \,  \omega_I \sim \ell^4$ \cite{ Firouzjahi:2025jja}.

The above was a summary of the main results for the RW bound states. Most of the above results apply to the polar and other spin field bound states as well.  The difference is only in the number of bound states. For example, for a given value of $\ell$, the polar sector has one extra mode compared to the axial sector. This additional mode is the algebraically special mode (ASM) with the spectrum,
\ba
\label{ASM-omega}
r_H\,  \omega_{\mathrm{asm}}= \frac{i}{6} (\ell-1) \ell(\ell+1)(\ell+2) \, .
\ea
In particular, ASM has the largest value of $|\omega|$ (the largest negative value of $\omega^2$) so it is the ground state in the polar sector.  On the other hand, all other $\ell-2$ modes in polar sector have the same energy spectrum as in axial sector. This establishes the isospectrality between the axial and polar sectors of the bound state solutions.  

Extending the above results to other spin fields, a field of spin $s$ has $\ell-s$ bound states for each value of $\ell$. They enjoy the same properties as the bound states in the axial sector. In particular, the approximate linear spectrum (\ref{omega-excited}) applies to all excited states independent of the spin or polarization sector. This universal  property of the spectrum will play key roles in our microstate counting.


\subsection{Minimum Length Scale and $\lmax$}

The above discussions were purely classical with no indications of quantum effects. Now, an important question is what is the minimum length scale which the above classical discussions are trusted? As shown in \cite{ Firouzjahi:2025jja, Firouzjahi:2025xjp}, the mode function $Z_{(p\ell m)}(r)$ for a given value of $\ell$ has $ \ell-p-2$ nodes in the RW sector. For the most excited states $p=1$, this gives   $\ell-3$ nodes. For $\ell\gg1$
the spacing between the nearby nodes  become very small. 
As we show in Appendix \ref{app-bound}, the wavelength associated to two nearby nodes in the spectrum of the most excited states 
 for $\ell \gg 1$ is given by,
\ba 
\label{lambda-min}
\lambda_\ell \simeq \frac{2 \pi r_H}{\ell} \, , \quad \quad \quad \quad   (\ell \gg 1) \, .
\ea
Physically, one expects that the above wavelength cannot be smaller than a minimum length scale $\lambda_{\mathrm{min}}$ below which the quantum effects into the background geometry cannot be ignored.  On the physical ground $\lambda_{\mathrm{min}}$ may be comparable to the length scale of quantum gravity $L_{\mathrm{Pl}}$, so we parametrize it as, $\lambda_{\mathrm{min}} =\gamma L_{\mathrm{Pl}}$ in which $L_{\mathrm{Pl}} = \sqrt{\hbar G}$ is the Planck length. A natural value of the parameter $\gamma$ may be simply $\gamma=1$, but we keep this free parameter undetermined. In this view $\gamma$ represents our ignorance as how to dynamically determine the minimum length scale where the quantum gravity effects become relevant. This uncertainty  is expected since our bound state analysis, as discussed above, are purely classical so their analysis by themselves cannot shed light on the minimum length scale relevant for the emergence of the quantum effects. 

From the above discussions we conclude that  there should be a maximum value of $\ell$, denoted by $\lmax$, beyond which the quantum gravity effects should enter into the analysis.  Using Eq. (\ref{lambda-min}), we obtain
\ba
\label{l-max}
\lmax = \frac{2\pi r_H}{\lambda_{\mathrm{min}}}  =
 c \frac{r_H}{L_{\mathrm{Pl}}}  \, ,\quad \quad \quad  \big(c\equiv \frac{2 \pi}{\gamma} \big) \, ,
\ea
where we have introduced the dimensionless constant $c$ for convenience.
A natural value with $\gamma=1$ is $c= 2 \pi$, but we keep the value of $c$ undetermined.   It is interesting that $\lmax$ scales like $r_H$ and this scaling is the key behind our final result for the share of the entropy of the bound states.

As explained before, in the interior of BH,  $r_*$ is timelike and $t$ is spacelike  so the spectrum $\omega$ is actually the momentum associated to $t$. Therefore, $|\omega|$ is not the energy. We postulate that in the semi-classical approach the energy spectrum of the massless bound state is related to the frequency $\omega$ as $E_i= \hbar |\omega_i | $ where $i$ is the general  labeling of the bound state. This would be a natural assumption for a quantized propagating field as in standard quantum field theory. However, the subtlety is that the bound state is not a propagating wave and its quantum nature is subtle. Having said this, we take the phenomenological approach and assume that the energy spectrum $E_i= \hbar |\omega_i | $ applies for the bond states similar to the propagating waves in the semi-classical limit.

The above was a review of the basic properties of the bound states solutions which will be used in our microstates counting and the entropy calculation. For further reviews on the properties of the bound state solutions see Appendix \ref{app-bound}. 


\section{Microstates Counting }
\label{micro-sec}

In this section we specify the microscopic degrees of freedom used in
the state counting problem. As mentioned previously, the basic one-particle state is labeled by $i=\{s,\ell,p,m\}$
where $(\ell, m)$ are the usual angular momentum labels  and $p$ labels the radial excitations. On the other hand, $s$ denotes the species sector and polarizations. For example, for 
gravitational perturbations $s=2$, we include both $\{\rm RW, \rm Z\}$ sectors. We denote the total number of massless species by $N$ but we do not know the exact value of $N$. Depending on the particle content and on which species may effectively be treated as light at the relevant scale, \(N\) can in principle be considerably larger than the two gravitational sectors. If only the gravitational perturbations intrinsically associated with the metric are retained, then \(N=2\).

We do not have a compact form  for $E_{s\ell pm}$ associated to the 
mode  $\{s,\ell,p,m\}$ but its values in two opposite limits of excited states $p=1,2,...,p_{\mathrm{max}}$ and low-lying states $p= \ell-2, \ell-3,...$ are known. For the excited states, 
\ba
E_{s\ell pm}=\hbar \omega_p\simeq \frac{p \hbar}{r_H}, \quad  p=1,2,..., 
\qquad \qquad (\rm {excited ~ states}) \, ,
\label{energy-excited}
\ea
while for the low-lying states they scale like $\ell^4$ for large $\ell$.

\subsection{Angular quantum numbers}

The  perturbation sectors begin at $\ell=s$, for example for tensor perturbations $\ell \ge 2$. 
We therefore take $ \ell=s,\ldots, \ell_{\max}$ with $\lmax$ given in Eq. (\ref{l-max}). Note that for a macroscopic 
BH, $\lmax$ is huge. For example, for a BH of solar mass, 
$\lmax \sim r_H/L_{\rm Pl} \sim \hbar^{-1} M_\odot G^{1/2} \sim 10^{38}$.  

For a fixed value of $\ell$, the azimuthal index  has the usual
range $m=-\ell,\ldots,\ell$.  Consequently, the degeneracy associated to $m$ is $2 \ell+1$. 
Correspondingly, the total angular degeneracy below the cutoff $\lmax$ is, 
\ba
N_{\rm ang}(\lmax)= 
\sum_{\ell=s}^{\lmax}(2\ell+1)\simeq  
(\lmax+1)^2 \simeq \lmax^2 \, .
\label{N-ang}
\ea
The fact that $N_{\rm ang}(\lmax) = \lmax^2$ plays important roles in the following analysis.  As we discussed before, the cutoff itself is parameterized as  $\lmax = c\,{r_H}/{L_{\mathrm{Pl}}}$
where $c$ is a dimensionless constant. Correspondingly, the leading angular degeneracy becomes $N_{\rm ang}(\lmax) \simeq c^2{r_H^2}/{L_{\mathrm{Pl}}^2}$.
Using the area formula $A=4\pi r_H^2$ we conclude  that
\begin{equation}
N_{\rm ang}(\lmax)
\propto
\frac{A}{L_{\mathrm{Pl}}^2}.
\label{N-ang-A}
\end{equation}
The important conclusion is that the  area scaling is already encoded in the angular state counting. However, the thermodynamic calculation is needed to determine the numerical coefficient multiplying this scaling.

\subsection{Radial excitation number}

For each fixed $(s,\ell,m)$, there are $\ell-s$ bound states. As mentioned above, except for the low-lying state (high $p$) and  the highly excited states (low $p$) we do not know the exact analytical form of  $E_{s\ell pm}$ as a function of $p$. However, as we shall see this does not prevent us  to obtain an analytic expression for the entropy. The reason is that the high-$p$ states are numerically unimportant when performing the saddle point approximation of the microcanonical analysis. More specifically, 
each state is accompanied by a thermal  weighting $e^{-\hbar \beta\omega_p}$
where $\beta$ is the auxiliary inverse temperature in the microcanonical saddle point analysis. Defining the dimensionless parameter $x\equiv{\hbar \beta}/{r_H}$, the analysis confirms that the saddle point occurs at $x=x_*  \gtrsim {\cal O} (1)$ 
so the thermal weights of high energy states are exponentially suppressed. Therefore, the leading contribution within each $\ell$ comes from the most excited states with $p= {\cal O}(1)$ with the associated weight $e^{- p x}$. In particular, the contribution of low-lying  and other modes with $p> p_{\mathrm{max}}$ are exponentially suppressed. To see this explicitly, let us look at the thermal weight of low-lying mode with $r_H \omega_I \sim \ell^4$
with the corresponding Boltzmann factor,
\begin{equation}
\exp(-\hbar \beta\omega) = \exp(-x\,r_H\omega)
\sim \exp(-x\ell^4) \, ,
\end{equation}
which is exponentially suppressed compared to $e^{- p x}$ contribution for low-$p$ modes. 

This is the key effect which allows us to obtain a compact analytical expression for the entropy of the bound states 
as we shall see in next section. However, it is important that all modes are present in the Hilbert space, only that the modes with high  values of $p$ are thermally suppressed in the state counting.

\section{Microcanonical Analysis} 
\label{micro-analysis}

In this section we construct the microcanonical density of states and derive the partition function used in the subsequent saddle-point calculation. This is motivated from the expectation that the total energy associated with the bound states are fixed so the microcanonical ensemble is the natural approach to start with.  In particular, there is no notion of Hawking temperature a priori. Alternatively, one can employ the canonical ensemble, assuming that the bound states are in thermal equilibrium in the interior of BH. We confirm that both ensembles yield to the same physical conclusions. In our microstate counting we do not consider the massive fields.  This is because they carry more energy than the massless states so 
their contribution  is thermally more suppressed (via the Boltzmann factor) 
compared to massless fields. 

\subsection{Microcanonical energy constraint} 

Let $i$ denotes a one-particle state, $i=\{s,\ell,p,m\}$  with the one-particle energy $E_i$. We have to specify the occupation number $n_i$ to proceed with the statistical descriptions. To prevent confusion, it is useful to distinguish carefully between the labels of a one-particle state and the occupation number of that state. The integer
$p$ specifies which member of the radial spectrum is being considered,
whereas the occupation number $n_i$ 
specifies how many quanta occupy that particular state. Thus, for
example, the states $\{s,\ell,p=1,m\}$ and $\{s,\ell,p=2,m\}$
are distinct one-particle states, while $n_{s\ell 1m}=0$ and
$n_{s\ell 1m}=1$ describe two different occupation configurations of
the state with $p=1$.

We assume that the total energy carried by the bound states is fixed, denoted by $E_{\rm bs}$. Since the black hole mass $M$ provides the natural macroscopic energy scale, it is convenient to characterize the chosen microcanonical energy of the bound states  by the dimensionless ratio,
\ba
f \equiv \frac{E_{\rm bs}}{M} \, ,
\ea
or equivalently $E_{\rm bs}=fM$. At this stage, $f$ simply parametrizes the microcanonical energy under consideration; we do not assume that it is a universal quantity or that it remains fixed when the black hole mass is varied. The fixed background approximation is best controlled in the 
small backreaction regime $E_{\rm bs}\ll M$, corresponding to $f\ll1$.

The microcanonical energy constraint is therefore given by
\ba
\label{E-micro}
\sum_i n_i E_i=E_{\rm bs}= f M \, ,
\ea
where, as mentioned before, $n_i$ denotes the occupation number of the mode $i$. There is also an independent scaling argument showing why it is natural to express $E_{\rm bs}$ as a dimensionless fraction of the black hole mass. The number of angular degrees of freedom up to the ultraviolet cutoff scales as $N_{\rm ang}\sim \ell_{\max}^2
\sim {r_H^2}/{L_{\rm Pl}^2}$. 
For the excited  bound states that dominate the relevant energy counting (with low value of $|\omega_I |$), the characteristic energy scale associated with an individual mode is $E_i\sim\hbar |\omega_i|\sim {\hbar}/{r_H}$. 
The collective energy scale associated with the angular degrees of freedom is therefore,
\ba
N_{\rm ang}E_i \sim \frac{r_H^2}{L_{\rm Pl}^2}
\frac{\hbar}{r_H}\sim M  \, .
\ea
Thus the characteristic total energy generated by the microscopic 
bound state degrees of freedom has the same scaling as the macroscopic BH mass. This supports the parametrization $E_{\rm bs}=fM$, while leaving the value of $f$ to specify the particular microcanonical energy realization. 

The next step is the prescription for the occupation number $n_i$. 
In our analysis below,  we assume the occupation number $n_i$ is a binary variable associated with the presence or absence of a distinct normalized  bound state mode. The  perturbation problem considered here is linear and yields a discrete set of independent bound state modes. Once a normalization convention is imposed, each mode $\psi_i$ represents a definite element of this discrete spectrum; an arbitrary rescaling of $\psi_i$ therefore does not constitute a new normalized spectral mode. We consequently regard a statistical configuration as a selection of a subset of the available bound state modes, with
\ba
n_i= \begin{cases}
0, & \text{if mode }i\text{ is not excited},\\
1, & \text{if mode }i\text{ is excited}.
\end{cases}
\ea
Thus $n_i$ is a mode-selection variable, not an amplitude variable. In particular, multiplying a normalized mode by a numerical factor does not represent a second independent excitation of that mode.  The total energy of a configuration is consequently written as,
\ba
E_{\rm bs}=\sum_i n_i E_i,
\qquad n_i\in{0,1}.
\ea

This binary prescription is appropriate to the linear perturbative regime in which the bound state modes constitute independent excitation channels. At nonlinear order, interactions among perturbations can generate mode coupling and invalidate the simple picture of independently switchable modes. Such nonlinear configurations are not included in the present construction. The choice $n_i=0,1$ should therefore be understood as a classical statistical description of configurations built from the discrete normalized bound state spectrum, rather than as a quantum-statistical exclusion principle like the Pauli exclusion principle. Having said this, one can verify that  the resulting entropy is found to be remarkably insensitive to relaxing this prescription: allowing multi-particle type excitations $n_i=0,1,2,...$ changes the  area-law coefficient only very weakly. This analysis was performed in detail in Appendix \ref{many-ni}. This robustness indicates that the leading $(A/\hbar G)$ scaling in our final result  is primarily associated with the number of distinct bound state modes, i.e. the degeneracy associated with the counting of $(\ell, m)$ sector captured by 
$N_{\rm ang}(\lmax)$,  while the precise treatment of repeated occupation of an individual mode is of secondary importance.

\subsection{Saddle Point Approximation}
\label{saddle}

The total energy is held fixed in which  $E[\{n_i\}] = \sum_i n_iE_i$. 
The microcanonical density of states is obtained by taking the trace of the energy constraint over the allowed configurations, following the standard quantum statistical mechanical prescription \cite{Kardar2007, Huang1987} $\Omega(E)=\operatorname{Tr}\delta(E-\hat H)$ where $\hat H $ is the Hamiltonian operator. Correspondingly, in energy basis, microcanonical density of states is given by, 
\ba
\Omega(E) = \sum_{\{n_i\}} \delta\Big(
E-\sum_i n_iE_i \Big).
\label{micro-Omega}
\ea
Here  the notation $\sum_{\{n_i\}}$ means $\sum_{\{n_i\}} \equiv \prod_i \sum_{n_i=0}^{1}$.

This expression contains the complete many-body Hilbert-space
counting. The lower limit $n_i=0$ should not be confused with the vacuum configuration, instead it means that a particular state may
be unoccupied while other states are occupied. Only when all
occupation numbers vanish, $n_i=0$ for $\forall i$, then the corresponding configuration corresponds to the vacuum. The vacuum  therefore represents  the background  Schwarzschild geometry where no perturbation is present. 

The entropy as usual  is obtained from the density of states according to,
\ba
S(E)=\ln\Omega(E)
\label{micro-entropy} \, .
\ea
The delta function in Eq.~\eqref{micro-Omega} can be
represented by an inverse Laplace transformation,
\ba
\delta(E-E')
=
\frac{1}{2\pi i}
\int_{\beta_0-i\infty}^{\beta_0+i\infty}
d\beta\,
e^{\beta(E-E')},
\label{delta-eq}
\ea
where the integration contour is a vertical line in the complex
plane with $\beta= \beta_0+ i y$ with $\beta_0 >0$ being fixed while 
$-\infty < y < +\infty$.

Substituting Eq.~\eqref{delta-eq} into
Eq.~\eqref{micro-Omega}, we obtain
\begin{align}
\label{Omega-int}
\Omega(E) &= \sum_{\{n_i\}} \frac{1}{2\pi i}
\int_{\gamma-i\infty}^{\gamma+i\infty}
d\beta\, \exp\Big( \beta E - \beta\sum_i n_iE_i \Big) 
\nonumber\\
&= \frac{1}{2\pi i} \int_{\beta_0-i\infty}^{\beta_0+i\infty} d\beta\,
e^{\beta E} \sum_{\{n_i\}} \exp\Big( -\beta\sum_i n_iE_i \Big).
\end{align}
Since the exponent is additive in the individual states, the
occupation-number sum factorizes, yielding,
\ba
 \label{factor-exp}
\sum_{\{n_i\}} \exp\Big( -\beta\sum_i n_iE_i \Big)
= \prod_i \sum_{n_i=0}^{1} e^{-\beta n_iE_i} =  \prod_i \big(1+ e^{-\beta E_i} \big) \, .
\ea
Consequently, defining the partition function $Z(\beta)$ via, 
\ba
Z(\beta) \equiv 
\prod_i
\big(1+ e^{-\beta E_i} \big) \, ,
\label{Z-micro}
\ea
the microcanonical density of states becomes,
\ba
\label{Omega-int2}
\Omega(E) = \frac{1}{2\pi i}
\int_{\beta_0-i\infty}^{\beta_0+i\infty}
d\beta\, e^{\beta E}Z(\beta) \equiv  \frac{1}{2\pi i}
\int_{\beta_0-i\infty}^{\beta_0+i\infty} d\beta\, e^{\Phi(\beta, E)} \, ,
\ea
in which,
\ba
\Phi(\beta, E)\equiv \beta E+ \ln Z(\beta) \, . 
\ea

Note that  the canonical partition function appears naturally as the
Laplace transform of the microcanonical density of states. In particular, note that $\beta$ is an integration variable which should not be confused with a 
given temperature as in canonical ensemble. We still have to perform the saddle point approximation while the
canonical partition function is an intermediate object in the inverse
Laplace representation.

Let $\beta=\beta_*$ represents the local extremum of $\Phi(\beta)$ so the saddle point satisfies $\Phi'(\beta_*)=0$, yielding
\ba
E(\beta_*) = - \frac{\partial \ln Z}{\partial \beta} \, .
\ea
Now, expanding to second order around the saddle point, we have,
\ba
\label{Phi-quad}
\Phi(\beta)= \Phi(\beta_*) + \frac{1}{2} \Phi''(\beta_*) (\beta - \beta_*)^2 +...
\ea 
To perform the saddle point integral, we consider a contour in the complex plane where $\beta= \beta_* + i y$ and perform the integral along the $y$ on the vertical axis. The integrand must be analytic in the region swept out by the deformation, apart from singularities that are not crossed, and the contributions from the connecting arcs must vanish. This yields,
\ba
\Omega(E)= \frac{1}{2\pi} \int_{-\infty}^{+\infty} dy e^{\Phi(\beta_*)} e^{-\frac{1}{2}  \Phi''(\beta_*) y^2} \, .
\ea
In order for the saddle point integral to be consistent we require $\Phi''(\beta_*) >0$ so the saddle point should be a minimum. Assuming this condition is met, we eventually obtain, 
\ba
\label{saddle1}
\Omega(E) \simeq \frac{e^{\Phi(\beta_*)}}{\sqrt{2 \pi \Phi''(\beta_*)}} \, .
\ea
Correspondingly, the entropy is calculated to be,
\ba
\label{S-eq1}
S\simeq \Phi(\beta_*) - \frac{1}{2} \ln \Big[ 2 \pi \Phi''(\beta_*) \Big]  \, .
\ea
The first term above is the leading term in entropy while the second term represents the subleading logarithmic corrections. In what follows, we consider only the leading term which yields the entropy associated to bound states as follows,
\ba
\label{S-eq2}
S(E) \simeq \beta_* E + \ln Z(\beta_*) \, .
\ea


\subsection{Explicit Partition Function }
\label{partition}

The bound state degrees of freedom are identified via $i=\{s,\ell,p,m\}$ so the partition function associated to the bound states in Eq. (\ref{Z-micro})
is cast into, 
\ba
\label{Z-full-eq}
Z_{\rm bs}(\beta) = \prod_{s} \prod_{\ell=s}^{\lmax} \prod_{m=-\ell}^{\ell}
\,  \prod_{p\in{\cal P}_{s,\ell}} (1+e^{-\hbar \beta\omega_{s\ell p}}).
\ea
Here ${\cal P}_{s,\ell}$ denotes the allowed radial spectrum in the
corresponding sector. For example, for the gravitational sector with $s=2$, 
\ba
{\cal P}_{\rm RW,\ell}
= \{1,\ldots,\ell-2\}, \qquad 
{\cal P}_{\rm Z,\ell} = \{1,\ldots,\ell-1\}.
\ea

Taking the logarithm and  taking into account the $2\ell+1$ degeneracy associated to the azimuthal index $m$  we obtain, 
\ba
\ln Z_{\rm bs} =  \sum_s  \sum_{\ell=s}^{\lmax}(2\ell+1)
\sum_{p=1}^{\ell-2} \ln\left(1+e^{- \hbar\beta \omega_{s \ell p}}\right) \, ,
\ea
where it is understood that $\omega_{s \ell p}$ means the value of $\omega_I$ with given values of $(s, \ell, p)$. 

So far everything in our mode counting was exact and there was no approximation. However, an important observation is that states with high values of $\omega_I$ are thermally suppressed and their contributions can be ignored. For example, for the low-lying modes with large values 
of $p$, we obtain 
\ba
e^{- \hbar\beta \omega_{s \ell p}} \, 
\sim  e^{- x \ell^4} \, , \quad \quad x \equiv \frac{\hbar \beta}{r_H} 
\ea
so their contribution in $\ln Z$ are exponentially suppressed. Correspondingly, for each sector $s$, only the modes with low value of $p$
have the dominant contributions in $\ln Z$. However, these modes satisfy the universal linear relation (\ref{omega-excited}) independent of $s$. Therefore,  noting that we have a total $N$ massless species, we obtain, 
\ba
\ln Z_{\rm bs}  &\simeq& N \sum_{\ell=2}^{\lmax} (2\ell+1) \sum_{p=1}^{p_{\rm max}}
\ln \left( 1+e^{-\hbar \beta\omega_{\ell p}} \right), \nonumber\\
&=& N \sum_{\ell=2}^{\lmax} (2\ell+1) \sum_{p=1}^{p_{\rm max}}
\ln \big( 1+ e^{- p x} \big) \, ,
\ea
where $p_{\rm max}$ represents the maximum value of $p$ beyond which the linear relation  (\ref{omega-excited}) is violated. 

The important conclusion is that the contributions of excited states are universal (independent of spin) with the thermal weight  $\ln \big(1+ e^{- p x} \big)$ in the partition function.  For $x >1$, this factor is approximated as $e^{- px}$. As a result the contribution of large $p$  falls off rapidly  and the dominant contribution comes from $p= {\cal O}(1)$ while the contribution of 
 modes with $p \sim  p_{\rm {max}} \sim \sqrt \ell$ are  exponentially suppressed. As a result, we can safely replace  $\sum_{p=1}^{p_{\rm max}} \rightarrow \sum_{p=1}^{\infty}$. 

With  the above approximation in mind,  we finally obtain,
\ba
\label{lnZ-eq1}
\ln Z_{\rm bs}\simeq  N  N_{\rm{ang}} \sum_{p=1}^{\infty} \ln \big( 1+ e^{- p x} \big)\, ,
\ea 
in which, $N_{\rm{ang}}$ is the angular degeneracy factor given in 
Eq. (\ref{N-ang}) scaling like $\lmax^2$.  For later convenience, let us define
\ba
\label{K-def}
K(x)\equiv \sum_{p=1}^{\infty} \ln \big( 1+ e^{- p x} \big) \, ,
\ea
so we have, 
\ba
\label{lnZ-eq2}
\ln Z_{\rm bs}(x) =  N \lmax^2 K(x) \, .
\ea

We comment that because of our particular choice of occupation number $n_i=0, 1$, the partition function exactly looks like a fermionic partition function with the Fermi function as in Eq. (\ref{K-def}). However, it is important to note that our selection analysis is classical and we do not assume quantum criteria like the Pauli exclusion principle for fermionic states.

Now we can examine the condition of our saddle point Gaussian integral, requiring $\Phi''(\beta_*) >0$. At the saddle point $\beta_*$ we have,
\ba
\Phi''(\beta)= \frac{\partial^2}{\partial \beta^2} \ln Z(\beta) = 
\frac{N \hbar^2 \lmax^2}{r_H^2} \frac{\dd^2 K(x)}{ \dd x^2} \, .
\ea 
On the other hand, one can easily check,
\ba
\frac{\dd^2 K(x)}{ \dd x^2} = \sum_{p=1}^{\infty} \frac{p^2 e^{-px}}{(1+ e^{-p x})^2} >0 \, .
\ea
As a result, $\Phi''(\beta_*) >0$ and our saddle point Gaussian integral was valid.


\subsection{Energy Constraint and the Entropy}
\label{energy}

Now we impose the fundamental constraint in the microcanonical ensemble that the energy $E_{\rm bs}$ carried by the bound states is fixed. From the saddle point condition we have
\ba
E_{\rm bs}= -\frac{\partial }{\partial \beta} \ln Z_{\rm bs}(\beta) = -\frac{\hbar}{r_H}
\frac{\partial }{\partial x} \ln Z_{\rm bs}(x) \, ,
\ea
where the derivative is evaluated at the saddle point
$\beta=\beta_*=r_H x_*/\hbar$.

Using the specific form of $Z(x)$ given in Eq. (\ref{lnZ-eq2}), we obtain
\ba
E_{\rm bs}= -\frac{\hbar N \lmax^2}{r_H} K'(x_*)=
\frac{\hbar N \lmax^2}{r_H} \sum_{p=1}^{\infty} \frac{p}{e^{p x_*}+1} \, .
\ea
For later convenience it is useful to define,
\ba
\label{F(x)}
F(x)\equiv -K'(x)=\sum_{p=1}^{\infty} \frac{p}{e^{p x}+1} \, .
\ea
In terms of this function, the energy associated with a saddle point $x_*$ is
\ba
\label{E-F(x)}
E_{\rm bs}(x_*)=\frac{\hbar N \lmax^2}{r_H}F(x_*) \, .
\ea
As mentioned before, we parametrized the microcanonical energy constraint  by the dimensionless ratio  $f\equiv {E_{\rm bs}}/{M}$. 
Correspondingly, using  the cutoff Eq. (\ref{l-max}), Eq. (\ref{E-F(x)}) gives
\ba
\label{f-F-eq}
f=2Nc^2F(x_*) \, ,
\ea
or equivalently
\ba
\label{F-c-eq}
F(x_*)=\frac{f}{2Nc^2} \, .
\ea
Since $F(x)$ is a monotonically decreasing function, specifying the microcanonical energy, or equivalently $f$, determines a unique value of the saddle point $x_*$. Conversely, a given value of $x_*$ selects the corresponding energy fraction through Eq. (\ref{f-F-eq}).

Having obtained the saddle point from Eq. (\ref{F-c-eq}), the entropy associated with the bound states is obtained from Eq. (\ref{S-eq2}), yielding, 
\ba
S_{\rm bs}=\beta_*E_{\rm bs}+\ln Z_{\rm bs}(\beta_*)
= N\lmax^2\left(x_*F(x_*)+K(x_*)\right) \, .
\ea
Now, using Eq. (\ref{f-F-eq}) and the relation (\ref{l-max}) between $\lmax$ and $r_H$, this can equivalently be written as,
\ba
\label{S-eq3}
S_{\rm bs}=\frac{f r_H^2}{2\hbar G}
\left(x_*+\frac{K(x_*)}{F(x_*)}\right) \, .
\ea
This result shows explicitly that the entropy scales with the horizon area. Indeed, using $A=4\pi r_H^2$, we finally obtain,
\ba
\label{S-final}
S_{\rm bs}=\alpha\frac{A}{\hbar G} \, ,
\ea
where the dimensionless parameter $\alpha$ is given by the saddle point via, 
\ba
\label{alpha-val}
\alpha(f,N,c)\equiv\frac{f}{8\pi}
\Big(x_*+\frac{K(x_*)}{F(x_*)}\Big) \, .
\ea
Equivalently, eliminating $f$ in favor of $x_*$ by means of Eq. (\ref{f-F-eq}), the coefficient may be written as
\ba
\label{alpha-x}
\alpha(x_*,N,c)=\frac{Nc^2}{4\pi}
\Big(x_*F(x_*)+K(x_*)\Big) \, .
\ea

The entropy associated with the bound states therefore scales with the area of the black hole. The key ingredient behind this scaling is the relation (\ref{l-max}), which relates $\lmax$ linearly to the black hole radius. This is also where the quantum gravity scale enters the analysis through the minimum length $\lambda_{\rm min}=\gamma L_{\rm Pl}$. The area law does not originate from a conventional thermal density of states. Rather, it originates from the angular degeneracy of the interior bound states together with a microscopic cutoff whose corresponding maximum angular momentum grows linearly with the horizon radius.

Let us denote the total entropy of the black hole from the Bekenstein-Hawking formula by $S_{\rm BH}$. Then from Eq. (\ref{S-final}) we obtain
\ba
\label{Sbs-SBH}
S_{\rm bs}=4\alpha S_{\rm BH} \, .
\ea
In order for the bound state to furnish the entire Bekenstein-Hawking entropy, we need $\alpha=1/4$. 

We next determine the thermodynamic temperature associated with the microcanonical ensemble. Before selecting a particular energy for bound states, the entropy should be regarded as a function $S_{\rm bs}(E_{\rm bs};M)$, where $M$ specifies the Schwarzschild background and therefore the bound state spectrum. The intrinsic statistical temperature is defined by
\ba
\label{Tstar-def}
\frac{1}{T_*}\equiv
\left(\frac{\partial S_{\rm bs}}{\partial E_{\rm bs}}\right)_M \, .
\ea
Using Eq. (\ref{S-eq2}), this yields  $\Big(\frac{\partial S_{\rm bs}}{\partial E_{\rm bs}}\Big)_M=\beta_*$ and consequently,
\ba
\label{Tstar}
T_*=\frac{1}{\beta_*}=\frac{\hbar}{r_H x_*} \, .
\ea
The temperature $T_*$ is thus an intrinsic statistical quantity obtained directly from the microcanonical construction. The derivative in Eq. (\ref{Tstar-def}) is taken at the fixed Schwarzschild background; the spectrum and the cutoff are held fixed while the microcanonical energy is varied. This is consistent with the semiclassical fixed background treatment adopted here when the energy stored in the bound state sector is sufficiently small that its gravitational backreaction can be neglected. 



\section{Physical Implications}
\label{implications}

The microcanonical analysis of the previous section describes a 
one parameter family of bound state configurations. For given values of $N$ and $c$, one may specify the energy fraction $f=E_{\rm bs}/M$, which determines the saddle point $x_*$ through Eq. (\ref{f-F-eq}), and hence determines both the entropy and the intrinsic statistical temperature $T_*$. Equivalently, one may use $x_*$ itself to parametrize this family, in which case $f(x_*)=2Nc^2F(x_*)$. 
No relation between $T_*$ and the Hawking temperature has been assumed so far. In this section we first examine the general microcanonical family and then consider the distinguished state selected by thermal matching with the Hawking temperature.

\subsection{Estimation of $\alpha$}

Our entropy formula for bound states Eq. (\ref{S-final}) has the desired property that it scales with the area. The parameter $\alpha$ is determined as a function of the parameters characterizing a given microcanonical state, $\alpha=\alpha(f,N,c)$. Here we estimate $\alpha$ in the regime where the saddle point is at large $x_*$. 

Let us define the ratio $\xi \equiv {f}/{2 N c^2}$
so the energy constraint equation (\ref{F-c-eq}) is translated into $F(x_*)=\xi$. We are particularly interested in the regime $\xi\ll1$, which includes the small backreaction regime $f\ll1$ for moderate or large values of $Nc^2$. In this limit the saddle-point equation admits a solution with $x_*>1$, so that an expansion in $q\equiv e^{-x_*}$ is well justified. Expanding $F(x_*)$ to second order, we obtain 
 \ba
 F(x_*) \simeq \frac{1}{1+ q^{-1}} + \frac{2}{1+ q^{-2}} \simeq q+ q^2 \, .
 \ea
Correspondingly, the energy constraint equation (\ref{F-c-eq}) 
becomes $q+ q^2 \simeq \xi$.  Solving this equation to quadratic order in $\xi$ yields $q \simeq \xi - \xi^2$. With this value of $q$, we obtain,
 \ba
 \label{xs-sol}
 x_*\simeq \ln\left(\frac{1}{\xi}\right)+\xi
 =\ln\left(\frac{2Nc^2}{f}\right)+\frac{f}{2Nc^2}
 +{\cal O}\left(\frac{f^2}{N^2c^4}\right) \, .
 \ea
Thus the assumption $x_*>1$ is self consistent for $\xi\ll1$.

Plugging the above value of $x_*$ into the function $K(x_*)$ and expanding to second order, we obtain,
\ba
K(x_*) \simeq q+\frac{q^2}{2}
\simeq \xi-\frac{\xi^2}{2}+{\cal O}(\xi^3) \, .
\ea
Finally, substituting the obtained values of $K(x_*)$ and $F(x_*)$ into our expression for $\alpha$ in Eq. (\ref{alpha-val}) yields,
\ba
\label{alpha-app}
\alpha \simeq \frac{f}{8\pi}\left[
1+\ln\Big(\frac{2Nc^2}{f}\Big)
+\frac{f}{4Nc^2}\right] \, .
\ea
The coefficient $\alpha$ therefore contains an overall factor of $f$, together with a logarithmic dependence on the ratio $Nc^2/f$. The latter can be sizeable when $Nc^2/f$ is large and can significantly modify a purely linear dependence on $f$.

Alternatively, in the limit $Nc^2/f\gg1$ we can solve for the combination $Nc^2$ from Eq. (\ref{alpha-app}), obtaining
\ba
\label{Nc2-eq}
Nc^2 \simeq \frac{f}{2}\exp\Big(\frac{8\pi\alpha}{f}-1\Big) \, .
\ea
This relation indicates that with given values of $\alpha$ and $f$, the cutoff $c$ scales like $1\sqrt N$. 

Here we provide some numerical examples. As a first example, suppose $c=2\pi$, $N=100$ and $f=0.1$. With these values, the full numerical solution of the saddle point equation yields $x_*\simeq11.277$ and $\alpha(x_*)\simeq0.049$. Interestingly, the corresponding values obtained from the analytic expressions (\ref{xs-sol}) and (\ref{alpha-app}) agree with the full numerical results to better than $10^{-7}$. As the second example, keeping $N$ and $f$ fixed but increasing the cutoff to $c=10$, we obtain $x_*\simeq12.201$ and $\alpha(x_*)\simeq0.053$, again with agreement better than $10^{-7}$ between the analytic approximation and the full numerical result.

Using the intrinsic statistical temperature derived in Eq. (\ref{Tstar}), its ratio to the Hawking temperature $T_H= \hbar/4 \pi r_H$ is,
\ba
\label{Tstar-TH-ratio}
\frac{T_*}{T_H}=\frac{4\pi}{x_*} \, .
\ea
Thus the two examples above give respectively $T_*/T_H\simeq1.114$ and $T_*/T_H\simeq1.030$. These examples represent generic members of the microcanonical family and no thermal matching with the Hawking temperature has been imposed.

\subsection{Thermal Matching with the Hawking Temperature}
\label{T=TH}

The microcanonical construction determines an intrinsic statistical temperature $T_*$ associated to the bound states.  A natural question is whether there is a distinguished member of this family for which the bound state sector is in thermal equilibrium with the black hole. We impose this condition by requiring $T_*=T_H$.  Using Eq. (\ref{Tstar-TH-ratio}), we concluded that  the thermal matching condition fixes the saddle point uniquely to $x_*=4\pi$. 

An important consequence is that the energy fraction is no longer an independent input for the thermally matched state. Instead, Eq. (\ref{f-F-eq}) predicts
\ba
\label{f-eq}
f_{\rm eq}\equiv\frac{E_{\rm bs}}{M}
=2Nc^2F(4\pi) \, ,
\ea
where the subscript ``${\rm eq}$" means in thermal equilibrium with the Hawking temperature. 
Similarly, using Eq. (\ref{alpha-x}), the entropy coefficient is predicted to be
\ba
\label{alpha-eq}
\alpha_{\rm eq}
=\frac{Nc^2}{4\pi}
\Big(4\pi F(4\pi)+K(4\pi)\Big) \, .
\ea
Thus, once $N$ and the cutoff parameter $c$ are specified, both the energy and entropy carried by the thermally matched bound state sector are determined by the spectrum.

It is useful to compare the entropy and energy fractions directly. From Eqs. (\ref{f-F-eq}) and (\ref{alpha-x}), for an arbitrary member of the microcanonical family we have
\ba
\label{share-general}
\frac{S_{\rm bs}}{S_{\rm BH}}
=\frac{1}{2\pi} \Big(x_*+\frac{K(x_*)}{F(x_*)}\Big)  f(x_*) \, .
\ea
At the Hawking thermal matching, $x_*=4\pi$, this becomes
\ba
\label{share-H}
\frac{S_{\rm bs}}{S_{\rm BH}}
= {\cal C}_{\rm {eq}} f_{\rm {eq}} \, , \qquad \qquad 
{\cal C}_{\rm {eq}} \equiv  2+\frac{K(4\pi)}{2\pi F(4\pi)} \, .
\ea
Since $e^{-4\pi}\ll1$, both $F(4\pi)$ and $K(4\pi)$ are dominated by the $p=1$ contribution and satisfy $K(4\pi)/F(4\pi)\simeq1$ so ${\cal C}_{\rm {eq}} \simeq 2.16$.

An important consequence of Eq. (\ref{share-H}) is the emergence of a universal relation between the energy and entropy fractions carried by the bound states in thermal equilibrium with the black hole. In particular, the ratio \((S_{\rm bs}/S_{\rm BH})\simeq 2.16 f_{\rm{eq}}\) is independent of the number of species \(N\) and the angular cutoff parameter \(c\). The leading contribution 2 follows from the Schwarzschild Smarr relation \(T_HS_{\rm BH}=M/2\), while the additional contribution originates from the microscopic partition function of the bound states. This relation is reminiscent of an equal sharing of energy and entropy, but with a nontrivial proportionality coefficient determined by the microscopic spectrum. Its universality reflects the common highly excited bound state spectrum of the massless species considered here, rather than a general thermodynamic identity.

For illustration, consider only the two gravitational sectors, $N=2$, and take the geometrically motivated value $c=2\pi$. Equations (\ref{f-eq}) and (\ref{alpha-eq}) then give $f_{\rm eq}\simeq5.5\times10^{-4}$ and 
$\alpha_{\rm eq}\simeq3.0\times10^{-4}$. 
Correspondingly, ${S_{\rm bs}}/{S_{\rm BH}}
=4\alpha_{\rm eq} \simeq1.2\times10^{-3}$.
The thermally populated gravitational bound states therefore carry only a small fraction of both the BH energy and entropy for these values of $N$ and $c$. At the same time, the small value of $f_{\rm eq}$ places this example safely within the regime where neglecting the backreaction of the bound states on the Schwarzschild geometry is expected to be a good approximation.

Now it is a good position to compare the canonical and microcanonical approaches. In our analysis, we have fixed the bound states energy and calculated the entropy and temperature using the microcanonical ensemble.
Alternatively, one could assume that the bound states are in thermal equilibrium with the BH and then look for their entropy and energy.  
The canonical mean energy at an arbitrary inverse temperature 
$\beta$ is $E_{\rm can}(\beta)=-\partial_\beta \ln Z$.  
This is precisely the saddle point equation of the microcanonical inverse Laplace transform. Therefore, when the canonical inverse temperature is chosen to equal the microcanonical saddle value, $\beta=\beta_*$, the two descriptions give the same energy and the same leading entropy,
$E_{\rm can}(\beta_*)=E_{\rm bs}$ and $S_{\rm can}(\beta_*)=S_{\rm bs}$,
up to the subleading fluctuation correction in Eq. (\ref{S-eq1}). The canonical and microcanonical ensembles thus provide two equivalent parameterizations of the same leading thermodynamic family. The former specifies $\beta$ and determines $E$, while the latter specifies $E$ and determines $\beta_*$. The Hawking condition $\beta_*=\beta_H$ selects the particular state given in Eqs. (\ref{f-eq}) and (\ref{alpha-eq}).

\subsection{Saturation of the Bekenstein-Hawking Entropy}
\label{extreme}

The thermally matched state discussed above provides only a small fraction of the Bekenstein--Hawking entropy for moderate values of $N$ and $c$. It is nevertheless interesting to ask under what conditions the bound state sector could account for the entire black hole entropy. In the present framework this corresponds to imposing $\alpha_{\rm eq}=1/4$ 
while maintaining thermal matching with the Hawking temperature, so that $x_*=4\pi$.

Using Eq. (\ref{alpha-eq}), the saturation condition becomes, 
\ba
\label{Nc2-sat}
Nc^2=\frac{1}{2 F(4\pi) {\cal C}_{\rm {eq}}}  \simeq 6.6\times10^4 \, .
\ea
Thus saturation requires a considerably larger value of the combination $Nc^2$ than in the perturbative examples considered in previous examples. 
Conversely, if we keep the geometrically motivated cutoff $c=2\pi$, Eq. (\ref{Nc2-sat}) requires $N\simeq 1.68\times10^3$. 
Therefore, with only the two gravitational sectors, $N=2$, the value $c=2\pi$ is far from sufficient to saturate the Bekenstein--Hawking entropy at the Hawking temperature. Increasing either the number of contributing species or the cutoff parameter increases the number of available bound state degrees of freedom and can formally bring the entropy to the Bekenstein--Hawking value.

There is, however, an important limitation. Once thermal matching and entropy saturation are imposed simultaneously, the corresponding energy fraction is no longer adjustable. Combining Eqs. (\ref{f-eq}) and (\ref{Nc2-sat}), we obtain
\ba
\label{f-sat}
f_{\rm sat} =\frac{1}{{\cal C}_{\rm {eq}}}  \simeq 0.463 \, .
\ea
Remarkably, this value is independent of $N$ and $c$. The parameters $N$ and $c$ determine whether the bound state spectrum contains enough states to reach $S_{\rm BH}$, whereas the energy fraction of the saturating thermally matched state is fixed by the spectral functions evaluated at $x_*=4\pi$. Thus, the equilibrium entropy fraction can equivalently be expressed as \(S_{\rm bs}/S_{\rm BH}=f_{\rm eq}/f_{\rm sat}\), establishing a direct connection between the universal energy-entropy relation and the formal saturation limit.

The value $f_{\rm sat}\simeq0.46$ is of order unity and therefore lies outside the controlled small backreaction regime $f\ll1$ used in treating the bound states as perturbations on a fixed Schwarzschild geometry. Consequently, Eq. (\ref{Nc2-sat}) should be viewed as a formal extrapolation of the statistical counting rather than as a controlled demonstration that the perturbative bound states alone reproduce the complete Bekenstein--Hawking entropy. A self consistent analysis including the gravitational backreaction of the bound state energy would be required to determine whether the saturation suggested by the counting survives in this regime.


\section{Summary and Discussions}
\label{summary-sec}

In this work we have considered the bound states perturbations inside the BH and looked at their contributions into the  Bekenstein-Hawking entropy. These bound states are defined as  perturbations which are regular at the center while their profiles fall off exponentially on the horizon. The bound state perturbations  are labeled by integers $(\ell, m, p)$ in which $(\ell, m)$ are the usual angular momentum labels while  $p$ represents the radial excitations.

The key property of the bound state perturbation is that, independent of the spin of the field,   the  excited states with $p=1,2,..., p_{\rm max}$ have a linear spectrum as in Eq. (\ref{omega-excited}). This is important as it relates the energy scale of these modes with the inverse radius of BH, $1/r_H$. Another important effect was that there is a maximum value for the angular momentum index $\ell$ beyond which the classical analysis of the bound states are not trusted. The cutoff  $\lmax$ is estimated by demanding that the smallest scale probed by the highly excited mode, $\lambda_{\rm min}$, should be larger than the length scale of quantum gravity. The minimum wavelength $\lambda_{\rm min}$ itself is determined by the nearby roots of the highly oscillatory modes (i.e. twice the distance between two roots),  yielding Eq. (\ref{l-max}). This is the key step in our analysis as this expression  relates $\lmax$ to $r_H$. Noting that the angular momentum degeneracy scales as   $N_{\rm ang} \propto \lmax^2$, we obtain the area law for  $N_{\rm ang}(\lmax)$ and eventually the area law for the entropy. 

We have calculated the entropy associated with the bound states  using the microcanonical ensemble approach. The total energy $E_{\rm bs}$ of the bound states is held fixed, and we have parametrized the corresponding microcanonical energy shell by the fraction $f=E_{\rm bs}/M$.  In order to neglect the backreaction on the background geometry, we are particularly interested in the regime $f \ll1$.  We have considered $N$ species of massless bound state perturbations in which all of them share the universal 
relation Eq. (\ref{omega-excited}) for their excited states.    At the linear perturbation regime, each bound state corresponds to one excitation so the mode counting involves the binary $n_i=0, 1$. If $n_i=0$ it means that the corresponding mode is not excited. On the other hand, $n_i=1$ means that the corresponding mode is turned on. Since the bound states are normalizable, then with appropriate normalization condition, a bound state  $i = \{ s,\ell, m, p\}$ represents a single copy of that spectrum. Having said this, we have demonstrated in Appendix \ref{many-ni} that our area scaling for the entropy of bound states is independent of the choice of occupation number and choosing a different set of occupation number changes 
only the numerical parameter $\alpha$  by a 
negligible  amount. 

We have established  the area law  $S_{\rm bs}= \alpha A/\hbar G$  in which the parameter $\alpha$ is determined by the parameters of the model and by the chosen microcanonical energy fraction. An important point in the present analysis is that the saddle point of the microcanonical counting itself defines an intrinsic statistical temperature $T_*$. Thus, for fixed $N$ and $c$, different choices of $f$ correspond to different members of a family of microcanonical states with different values of $T_*$. The Hawking temperature is not imposed in constructing this family. Instead, requiring thermal equilibrium with the BH selects the particular state for which $T_*=T_H$. For this state the saddle point is fixed to $x_*=4\pi$, and the energy fraction $f$ is no longer a free input but is determined by $N$ and $c$. The entropy share is also fixed and is proportional to the energy share 
by a universal constant independent of parameters such as $N$ and $c$. 
For example, considering only the two gravitational sectors with $N=2$ and taking $c=2\pi$, we find $f_{\rm eq}\simeq5.5\times10^{-4}$ and $\alpha_{\rm eq}\simeq3.0\times10^{-4}$. The small value of $f_{\rm eq}$ also means that this example lies safely within the regime in which the backreaction on the Schwarzschild geometry can be neglected.
It is also interesting to ask whether the bound states can furnish the entire Bekenstein--Hawking entropy. At Hawking equilibrium this requires a much larger number of available states, corresponding to $Nc^2\simeq6.6\times10^4$. 
More importantly, once both thermal equilibrium and $\alpha=1/4$ are imposed, the bound states carry an energy fraction $f\simeq0.46$, independently of the particular values of $N$ and $c$. This is no longer in the small backreaction regime. Therefore, while the area scaling of the entropy follows within the controlled perturbative calculation, the possibility that the bound states account for the full Bekenstein--Hawking entropy should be regarded as an extrapolation which requires a consistent treatment of their gravitational backreaction.

There are a number of directions in which the current line of research can be extended or critically examined. The microstates associated with the bound states are the heart of our proposal so it would be helpful to understand the physical meaning of these bound states in more details. For example, our analysis so far were mainly semi-classical, imposing the assumption $E_i= \hbar |\omega_i|$ in analogy with propagating perturbations.  It would be helpful if one looks at their quantum behaviours by quantizing these mode functions. This may also shed some light on  the natural value of the minimum length cutoff $\lambda_{\rm min}$ and correspondingly the value of $c$. Second, one can extend the current analysis to higher dimensional spherically symmetric BHs and examine whether the general features of the bound state solutions and their contributions in entropy follow the same pattern as in four-dimensional Schwarzschild BH. Similarly, one may extend the formalism of bound states perturbations to  the Kerr and charged BHs, which have double horizons, 
and see if the results  for their entropy match with the properties obtained here for the Schwarzschild BH. 

\vspace{1cm}

{\bf Acknowledgments:}  
We thank Masoud Molaei, Shinji Mukohyama, Kazem Rezazadeh  and Ali Seraj  for useful discussions and correspondences. The author acknowledges the use of Maple software package for analytical and numerical analysis.  This work is supported by the INSF  of Iran under the grant  number 4045105. \\

The author acknowledges the assistance of OpenAI's ChatGPT (GPT-6)  in exploring mathematical calculations, discussing theoretical aspects of the analysis and editing the manuscript. The physical motivation and technical formulation of the analysis were developed by the author. All results were  independently and critically examined by the author who assumes full responsibility for the scientific content of the manuscript.

\appendix

\section{Unrestricted Occupation Numbers}
\label{many-ni}

As we discussed in section \ref{micro-analysis}, our analysis for the microstates counting was based on binary counting with $n_i=0,1$.  Since the system is linear, the modes are uncorrelated and a given mode can be either on or off, independent of other modes. In addition, the bound states are normalized so there is no freedom in the amplitude of the mode. 
This motivates us to consider  each normalized mode $i$ as a single realization  in bound state solutions. This justifies the binary choice $n_i=0,1$. Having said this, one may worry that our area law is an artifact of the above binary variables. Here we demonstrate that our area law results are quite insensitive to the binary or any other occupation number selection.

To see this, let us consider the extreme possibilities that all modes can be occupied unlimitedly, corresponding to $n_i=0,1,2,3,...$. Most of our microcanonical analysis in sections \ref{saddle}, \ref{partition} and $\ref{energy}$ formally go through. However, the explicit form of the partition function $Z(\beta)$ is now modified. More specifically, since now we consider the unrestricted occupation number, then we expect that the partition function takes the form of the Bose-Einstein distribution for the bosonic fields. More specifically, the density of  states is the same as in Eq. (\ref{micro-Omega}) but the occupation number modifies the summation $\sum_{\{n_i\}}$ into,  
\ba
\sum_{\{n_i\}} \equiv \prod_i
\sum_{n_i=0}^{\infty}.
\label{occup-sum2}
\ea
After implementing this modification, the term inside the complex plane integral in Eq. (\ref{factor-exp}) is modified into,
\ba
\sum_{\{n_i\}} \exp\Big( -\beta\sum_i n_iE_i \Big) =
\prod_i \sum_{n_i=0}^{\infty}
e^{-\beta n_iE_i} \, .
\label{factor-exp2}
\ea
With $\Re(\beta) = \beta_0 >0$, the above sum is a geometric series, yielding,
\ba
\sum_{n_i=0}^{\infty} e^{-\beta n_iE_i}
= \frac{1}{1-e^{-\beta E_i}} \, .
\ea
Consequently,  the partition function $Z(\beta)$ is modified into, 
\ba
Z(\beta) \equiv  \prod_i \frac{1}{1-e^{-\beta E_i}} \, .
\label{Z-micro2}
\ea
As expected, this has the form of the Bose-Einstein partition function for bosonic fields where there are no restrictions on $n_i$. 

The saddle point analysis goes formally unchanged, yielding,
\ba
S= \beta_* E + \ln Z(\beta_*) \, .
\ea
 
The explicit form of the partition function, after performing the mode counting over $i= \{s, \ell, m, p, \}$, is given by,  
\ba
\ln Z(x)= N \lmax^2 K_2(x) \, ,
\ea
in which now the function $K_2(x)$ is given by,
\ba
K_2(x)\equiv -\sum_{p=1}^{\infty} \ln \big( 1- e^{- p x} \big) \, .
\ea
The function  $K_2(x)$ should be compared with $K(x)$ defined in Eq. (\ref{K-def}). Indeed, one can easily check that for the saddle point $x_* = {\cal O}(1)$ the two functions $K(x)$ and $K_2(x)$ are very similar.  However, more importantly for our purpose, note that $\ln Z(x)$ has exactly the same dependence on $\lmax^2$ as before which is the main reason behind the entropy area law. 

The energy constraint equation is now given by $f=2 N c^2 F_2(x_*)$, 
or equivalently,
\ba
\label{F-c-eq2}
F_2(x_*) = \frac{f}{2 N c^2} \, ,
\ea
where, as in the main text, $f=E_{\rm bs}/M$ labels the chosen microcanonical energy shell. The new function $F_2(x)$ is given by,
\ba
\label{F(x)-2}
F_2(x)\equiv \sum_{p=1}^{\infty} \frac{p}{e^{p x} -1} \, .
\ea
Again, the new function $F_2(x)$ should be compared with the old 
function $F(x)$ given in Eq. (\ref{F(x)}).  For $x_* = {\cal O}(1)$ they look very similar while for $x\ll 1$ they behave  differently.

After imposing the energy constraint, the entropy at the saddle point is 
given by the same area law as in Eq. (\ref{S-final}) 
but now the dimensionless parameter $\alpha$ is given by, 
\ba
\label{alpha-val2}
\alpha(f,N,c) \equiv \frac{f}{8 \pi} \Big[ x_* + \frac{K_2(x_*)}{F_2(x_*)} \Big]  \, .
\ea
So we recover the area law and the only change is on the dependence of the prefactor $\alpha$ on $c$. The conclusion is that the area law is independent of the form of the occupation number $n_i$ and is mainly controlled by the angular degeneracy $N_{\rm{ang}}(\lmax)$. But   $N_{\rm{ang}}(\lmax)$ is just $\lmax^2 \propto A$ so the area law is independent of the occupation numbers.

One can also check numerically that the change in $\alpha$ is very small, typically at the order $10^{-4}$ or so.  It may look surprising that a binary counting and an unrestricted occupation counting yield such close results. Mathematically, however, there is no surprise. As we mentioned before, the Fermi functions $(K(x), F(x))$ and the Bose-Einstein functions $(K_2(x), F_2(x))$ look very similar for large enough values of $x$. In the regime relevant to our analysis $x_* \gtrsim 1$, and in particular for the large-$x_*$ saddles considered in the main text, the deviations between these two sets of functions are exponentially small, at the order 
$e^{-2 x_*} \ll 1$.

\section{Bound State Solutions Inside the BHs}
\label{app-bound}

In this section we review basic properties of the gravitational bound state solutions in the interior of BH, for further details see \cite{Firouzjahi:2025jja, Firouzjahi:2025xjp}. The main results can be extended to other spin fields as well. 

The bound states are perturbations in the interior of BH with the imaginary frequency $\omega^2 <0$. They are defined with the regular boundary condition at the center $r=0$ while their profiles fall off exponentially on the horizon so they do not leak to the exterior region.

The mode functions have the usual RW/Z type equations. Specifically, expanding the perturbations in spherical harmonic basis and decomposing the remaining part of the wave function via $Z(t, r)= e^{- i \omega t} Z(r)$, the mode function satisfies the usual RW/Z equation,
\ba
\label{RW-Z}
\frac{\dd^2}{\dd r_*^2} Z(r_*) + \Big(\omega^2 - V_{\ell} (r_*) \Big) Z(r_*) =0\, ,
\ea
in which $V_{\ell}(r_*)$ is the effective potential and 
$r_*$ is the tortoise coordinate defined as usual via Eq. (\ref{r-star})
with the solution given in Eq. (\ref{r-star-sol}). 
One key difference compared to the exterior region is that now $r_*$ is timelike in the interior while $t$ is spacelike. In addition, $-\infty < r_* < 0$ in which the horizon is mapped to $r_* \rightarrow -\infty$ while the center corresponds to $r_* \rightarrow 0$.

For the RW (axial) sector the effective potential  has the form,
\ba
\label{Veff_spin}
V^{-}_{\ell}  = \Big(1- \frac{r_H}{r} \Big)  \left[ \frac{\ell (\ell+1)}{r^2} - \frac{3 r_H}{r^3} \right] \, ,
\ea 
while for the Z (polar) sector, denoted by $V^{+}_{\ell}$,  it has a more complicated form \cite{Chandrasekhar1983, Firouzjahi:2025xjp}. As is well-known, $V_\ell^\pm$ are related to each other via a Darboux transformation \cite{Chandrasekhar1983} so they share the same spectrum for the quasi normal modes (QNMs) spectrum in the exterior region. While the corresponding Darboux transformation formally does hold in the interior region, but the existence of singularity at $r=0$ may prohibit the validity of the  Darboux transformation. However, as shown specifically in \cite{Firouzjahi:2025xjp}, the isospectrality between the polar and axial bound states does hold in the interior region as well. Specifically, for a given value of $\ell$ the axial perturbations have $\ell-2$ bound states which are shared by the polar sector as well. In addition, the polar sector has an extra mode, the algebraically special mode (ASM) which is not shared in axial sector.

The spectrum can be labelled by the radial number $p=1,2,..., \ell-1$, in which $p=\ell-1$ is the ground state associated to the ASM which exists only in the polar sector. The remaining $\ell-2$ modes have the equal spectrum in both sectors. The spectrum of ASM is given by,
\ba
\label{ASM-omega2}
r_H\, \omega_{\mathrm{asm}}= \frac{i}{6} (\ell-1) \ell(\ell+1)(\ell+2) \, ,
\ea 
so it scales like $\ell^4$ for $\ell \gg 1$.  All other spectrum have spectrum   smaller value than $| \omega_{\mathrm{ASM}}|$. In table \ref{table:omegaI} the numerical values of $\omega_I$ for the polar (Z) sector are presented for $\ell=2,3,..., 10$. The ASM are located on the diagonal part of the table while the high excited states are in top rows.

\begin{table*}
\centering
\scalebox{1}{
\begin{tabular}{|c||c|c|c|c|c|c|c|c|c|}
\hline
$p$ & $\ell=2$ & $\ell=3$ & $\ell=4$ & $\ell=5$ & $\ell=6$ & $\ell=7$ & $\ell=8$ & $\ell=9$ & $\ell=10$\tabularnewline
\hline
1 & 4 & 1.71 & 1.39 & 1.27 & 1.21 & 1.17 & 1.15&1.13&1.11 \tabularnewline
\hline
2 & $-$ & 20 & 5.75 & 3.97   &  3.32    &   2.98 &2.78&2.65&2.55\tabularnewline
\hline
3 & $-$ & $-$ & 60 &  13.77  & 8.39 & 6.49&5.54&4.99&4.63  \tabularnewline
\hline
4 & $-$ & $-$ & $-$ & 140 & 27.79 & 15.43&11.20&9.15&7.97\tabularnewline
\hline
5 & $-$ & $-$ & $-$ & $-$ & 280 &50.20&25.96&17.90&14.08\tabularnewline
\hline
6 & $-$ & $-$ & $-$ & $-$ & $-$ & 504&83.83&41.01&27.11\tabularnewline
\hline
7 & $-$ & $-$ & $-$ & $-$ & $-$ & -&840& 131.86&61.69\tabularnewline
\hline
8 & $-$ & $-$ & $-$ & $-$ & $-$ & -&-& 1320&197.90\tabularnewline
\hline
9 & $-$ & $-$ & $-$ & $-$ & $-$ & -&-& -&1980\tabularnewline
\hline
\end{tabular}
}
\vspace{0.5cm}
\caption{$r_H \omega_I$ for $\ell=2,..., 10$ in polar (Z) sector.   For a given value of $\ell$, there are $\ell-1$ bound states in which the state with $p=\ell-1$ is the ASM with the spectrum given in Eq. (\ref{ASM-omega}). The remaining $\ell-2$ spectrum are shared by the axial (RW) sector. To save space, the values of $\omega_I$ to two  significance digits are presented. 
}
\label{table:omegaI}
\vspace{.5cm}
\end{table*}

The spectrum of $\omega_I$ for modes other than ASM are not known analytically. However, for two opposite limits of excited states with 
$p=1,2,..., p_{\rm{max}}$ and the low lying states with $p=\ell-2, \ell-3,...$  their spectra are known to very good accuracies \cite{Firouzjahi:2025jja}. As we discussed in the main text, the ASM and other low-lying states are thermally irrelevant in the partition   function and only the excited states with relatively low values of $\omega_I$ are relevant in the microstate counting. Below, we review the analysis yielding to their spectrum.


\subsection{Spectrum of Highly Excited States}
\label{excited-states}

Before presenting the spectrum of the highly excited states for $\ell \gg1$, it is convenient to write the Schwarzschild metric in a new coordinate as follows. let us define the angular coordinate $\chi$ via \cite{Firouzjahi:2025jja}, 
\ba
\label{chi}
 r= r_H \cos(\chi)^2 \, ,  
\ea
in which  the metric in the new coordinate takes the following form,
\ba
\label{BH-metric-chi}
\dd s^2 = r_H^2 \cos^4(\chi) \big( - 4 \dd \chi^2 + \dd \Omega^2 \big)  + \tan^2(\chi) \dd t^2  \, .
\ea 
One can absorb the factor $4$ above via rescaling of $\chi$ to make the metric along the directions $(\chi, \theta, \phi)$ conformal, but is 
better to keep it as above where  the periodicity of $\chi$ takes a simple form. In this coordinate, the horizon  $r=r_H$ corresponds to $\chi=0$ while the  center is mapped to $\chi=\frac{\pi}{2}$. Curiously,  the periodic structure of $\tan(\chi)$ indicates that  $\chi$ maybe analytically 
extended to other parts of the whole manifold such as the white hole region as well. But here we only consider the interior of BH with $0< \chi < \frac{\pi}{2}$. 

As demonstrated in  \cite{Firouzjahi:2025jja} the solution for the highly excited states in the regions $\chi \rightarrow 0$ and $\chi \rightarrow \frac{\pi}{2}$ can be written in terms of the Bessel functions. In the region near the horizon, the regular solution has the following form,
\ba
\label{sol-1b}
A_-= \sqrt{\chi} \,  J_{2 \omega_I r_H} (\nu \chi) \, , \quad \quad (\chi \rightarrow 0) 
\ea  
in which the index $\nu$ in the large-$\ell$ limit is given by  $\nu \simeq 2 \ell $.
The index $\nu$ receives subleading corrections of the form $\omega^2/\ell^2$ which can be ignored in large $\ell$ limit. 

The regular solution near the center  has the following form, 
\ba
\label{sol-2b}
A_+ =   D_1 \sqrt{\chi-\frac{\pi}{2}} J_{4} \Big(\nu (\chi-\frac{\pi}{2}) \Big)\, ,
\ea
in which $D_1$ is a constant of integration which is fixed by matching the two solutions $A_-$ and $A_+$ at an intermediate point in the region $0< \chi < \frac{\pi}{2}$ where these two solutions have overlaps. 

The details of the matching are presented in \cite{Firouzjahi:2025jja} with the conclusion,
\ba
\label{high-excited}
\omega_I = \frac{p}{r_H} + {\cal O}(\ell^{-1} ) \, , \quad \quad (\ell \gg 1) \, ,
\ea
in which $p=1,2,..., p_{\rm {max}}$. With some efforts one can show that 
$p_{\rm {max}} = b \sqrt{\ell}$  beyond which the linear relation (\ref{high-excited}) shows significant deviation. The numerical factor $b$ is  $ {\cal O}(1)$.  

We have checked that the linear relation (\ref{high-excited}) is very accurate for $\ell \gg 1$ and for $p = {\cal O}(1)$. This can be seen in table {\ref{table-large-ell}} where the accuracy increase as $\ell$ increase. In our macroscopic BH with $\lmax \rightarrow \infty$, the linear 
relation  (\ref{high-excited}) is extremely accurate.

\begin{figure}[t]
\begin{center}
	\includegraphics[scale=0.49]{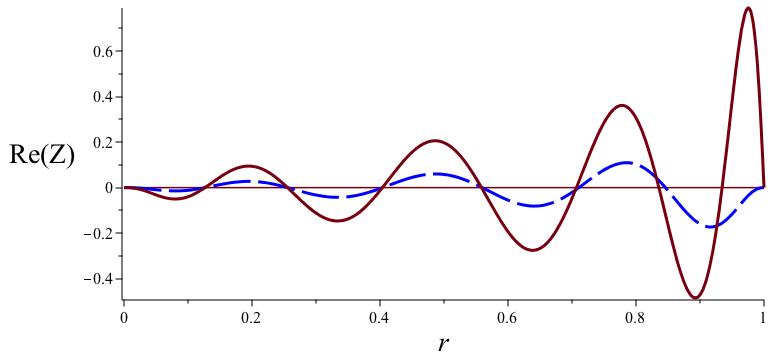}
	\end{center}
\caption{   The profile of  the axial (RW) sector with $\ell=10$ for the most excited state $p=1$ (solid red) and the next most excited state $p=2$ (dashed blue).  Each curve has $\ell-p-2$ nodes in the region $0< r<1$. We have set $r_H=1$ for convenience.  
\label{profile-excited}
\vspace{0.5cm}
}
\end{figure}

Now we estimate $\lmax$ and its relation to $\lambda_{\rm {min}}$. For $\ell \gg 1$, the mode function $Z(r)$ is highly oscillatory as can be seen from the structure of Bessel functions for the mode functions $A_-$ and $A_+$. In Fig. \ref{profile-excited} we have plotted the most excited state $p=1$ and the next most excited state $p=2$ for the axial (RW) sector with $\ell=10$. As seen, the mode function has $\ell-p-2$ nodes in the interval $0< r< r_H$. In particular, for $\ell \gg1$, the spacing between two nearby nodes become very small. On the physical ground, the mode function cannot probe a distance shorter than the quantum gravity scale $L_{\mathrm{Pl}}$. For wavelength smaller than 
$L_{\mathrm{Pl}}$ we cannot neglect the quantum fluctuations of the spacetime itself and our classical analysis yielding to the linear relation (\ref{high-excited}) is not valid. This in turn impose a cutoff for $\lmax$ above which the quantum gravity scales become relevant.

To estimate $\lmax$, a reasonable criteria is to demand that the wavelength associated to the most excited state with $p=1$ and $r_H \omega_I \rightarrow 1$ to be larger than quantum gravity scale. Using the large argument limit of the Bessel function,
\ba
J_\mu(x) \simeq \sqrt{\frac{2}{\pi x}} \cos\Big( x- \frac{\mu \pi}{2} - \frac{\pi}{4} \Big) \,,  \quad \quad ( x\gg 1) \, ,
\ea
we find that the wavelength  of $A_-$ near the horizon 
is approximately given by,
\ba
(2 \ell) \Delta \chi \simeq  2 \pi \, .
\ea
Here $\Delta \chi$, representing the angular wavelength, is twice the distance between two nearby roots. 
In the limit of interest $\ell \gg 1$ this yields $\Delta \chi \simeq \frac{\pi}{\ell}$.
Now we have to translate the angular distance $\Delta \chi$ to a proper length scale in the metric (\ref{BH-metric-chi}). Taking into account the line element  near the horizon $\chi=0$ the corresponding wavelength is given by,
\ba
\label{lambda}
\lambda \simeq (2 r_H) \Delta \chi = \frac{2 \pi r_H}{\ell} \, .
\ea 
Now our physical criteria is that the above wavelength should be larger than the quantum gravity scale. Taking into account the numerical uncertainty of whether $L_{\rm{Pl}}$ or a similar length scale can be used, we parametrize this via an undetermined parameter $\gamma$ as $\lambda_{\rm{min}} =  \gamma L_{\rm{Pl}}$ with the expectation that 
$\gamma = {\cal O}(1)$.  
Demanding that the wavelength (\ref{lambda}) be larger than this minimum length scale, $\lambda \ge \lambda_{\rm{min}}$,  imposes an upper bound on $\ell$ via,
\ba
\label{lmax-eq}
 \lmax=  c \frac{r_H}{L_{\rm{Pl}}}  \, , \quad \quad 
c\equiv \frac{2\pi}{\gamma} \, .
\ea
As explained in the main text, Eq. (\ref{lmax-eq}) is the key relation behind the BH area law in our approach. First, we see that $\lmax$ scales linearly with $r_H$ so the angular degeneracy $N_{\rm{ang}}(\lmax)$ scales like $\lmax^2 \propto A$ as given in Eq. (\ref{N-ang-A}).  Second, the quantum gravity effect appears in our analysis  via the cutoff $\lmax$ given in  Eq. (\ref{lmax-eq}). This is essential to provide a semi-quantum estimation of the bound state entropy area law in our approach.

\vspace{1cm}


\bibliography{bound-entropy}{}

\bibliographystyle{JHEP}

\end{document}